\documentclass[twocolumn,showpacs,preprintnumbers,amsmath,amssymb]{revtex4}

\usepackage{graphicx}
\usepackage{dcolumn}
\usepackage{bm}

\def\mbf{\mathbf}

\begin{document}

\preprint{APS/123-QED}

\title{
Nonequilibrium relaxation analysis 
of quasi one dimensional frustrated XY model \\
for charge density waves in ring crystal
}

\author{Tomoaki Nogawa}
\email{nogawa@statphys.sci.hokudai.ac.jp}

\author{Koji Nemoto}
\email{nemoto@statphys.sci.hokudai.ac.jp}

\affiliation{%
Division of Physics, Hokkaido University,
Sapporo, Hokkaido 060-0810 Japan
}%

\date{\today}

\begin{abstract}
We propose a model for charge density waves in ring shaped crystals, 
which depicts frustration between intra- and inter-chain couplings 
coming from cylindrical bending.
It is then mapped to a three dimensional uniformly frustrated XY model 
with one dimensional anisotropy in connectivity.
The nonequilibrium relaxation dynamics 
is investigated by Monte Carlo simulations 
to find a phase transition which is quite different 
from that of usual whisker crystal.
We also find that the low temperature state is 
a three dimensional phase vortex lattice 
with a two dimensional phase coherence in a cylindrical shell 
and the system shows power law relaxation in the ordered phase. 
\end{abstract}


\maketitle


\section{Introduction}

Transition metal chalcogenides such as NbSe$_3$ and TaS$_3$
are known as quasi one dimensional materials. 
Lowering temperature, they take a phase transition 
to the charge density wave (CDW) phase 
where both of atom positions and electronic charge density 
are modulated in twice of the Fermi wave number \cite{Gruner88}. 
Recently, Tanda and Tsuneta {\it et al.} have succeeded to synthesize 
various types of single crystals in closed loop shapes, 
e.g., simple ring, M\"obius ring and figure-of-eight, 
of these materials 
by the chemical vapor transportation method\cite{Tanda02,Tsuneta04}. 
X-ray diffraction measurement shows that they have 
the same crystalline structure with usual whisker crystals 
and the temperature dependence of conductivity indicates a CDW transition 
at the temperature slightly lower than the critical temperature 
of whisker crystals \cite{Tsuneta03}. 
The influence of the crystal geometry or topology 
on the property is an interesting problem.

Recently, Shimatake and Toda investigated 
nonequilibrium relaxation dynamics of NbSe$_3$ 
by using ultrafast laser \cite{Shimatake06}. 
They measured both of ring and whisker crystals with comparable dimensions 
and found a significant difference between them 
in the low temperature CDW phase.
The relaxation time for the initial rapid decay 
shows divergent behavior at the transition temperature 
for a whisker crystal but does not for a ring crystal. 
It suggests that the phase transition disappears 
or the type of a phase transition changes for ring crystals.
In this article, 
we propose a simple model for CDWs in ring crystals 
and analyze the nonequilibrium relaxation dynamics 
by Monte Carlo method to understand the reason 
for such difference.

\section{model for cdw in ring crystal}

\begin{figure}
\includegraphics[trim=60 615 70 -205,scale=0.37,clip]{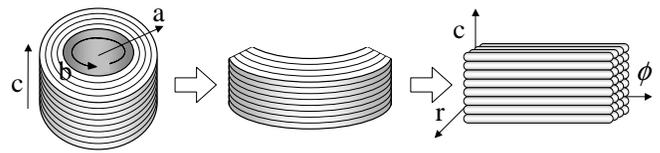}
\caption{\label{fig:ring}
Schematic diagram of ring crystal and mapping onto cuboid. 
}
\end{figure}

At first, we introduce a phase field model for CDW in a ring crystal. 
The crystal axes are named a-, b-, and c-axes 
as illustrated in Figure \ref{fig:ring}. 
The quasi one dimensional chains run along the b-axis.
The position in a sample is expressed as
$\mbf{r}=(x_a,x_b,x_c) = (r, r \phi, x_c)$ 
by using cylindrical coordinates.
The modulated part of charge density can be expressed as  
$\rho(\mbf{r}, t) = \rho_0 \cos \left(
Q(r) x_b + \theta(\mbf r, t) \right) $ 
in each chain. 
Here, $Q(r)$ is the mean wave number of a chain at radius $r$ 
and $\theta(\mbf r, t)$ is the phase fluctuation variable.
Periodic boundary condition along a chain yields 
$Q(r)2\pi r=2\pi N_w(r)$ where $N_w(r)$ is a number of waves in a chain. 
$Q(r)$ possibly deviates from the natural wave number of CDW, $Q_0=2k_F$, 
to synchronize the charge density modulations 
between neighboring chains. 
Such commensuration between the chains
neighboring on the radial direction results strain. 
The stress grows with the thickness 
and is released by making some imperfection, 
such as a jump of $N_w$, by every several chains. 
We consider that the characteristic relaxation dynamics 
of a ring crystal is due to such conflict 
between two requirements for the natural wave length inside each chain 
and for period matching between neighboring chains.

Here we suppose bundles of looped chains in which no defect exists
and then $N_w(r)$ is constant. 
The mean wave number decreases with radius as 
$Q(r) = N_w / r$ in each bundle. 
The wave number is larger than the  natural one 
for inner edges and smaller for outer edges. 
The Coulomb interaction $V_\alpha$ at the boundary 
of neighboring two bundles is 
proportional to the product of charge densities as  
\begin{eqnarray}
V_a 
\propto \int dx_b
\cos[Q_\downarrow x_b+\theta_\downarrow (x_b)]
\cos[(Q_\uparrow x_b+\theta_\uparrow (x_b)] 
\nonumber \\ 
= \frac{1}{2} \int dx_b
\{\cos[ \theta_\downarrow(x_b) - \theta_\uparrow(x_b) - \Delta Q x_b] 
\nonumber \\ 
+ \cos[ \theta_\downarrow(x_b) + \theta_\uparrow(x_b) 
+ (Q_\downarrow+Q_\uparrow) x_b]\}
\label{eq:coupling}
\end{eqnarray}
The down- $(\downarrow)$ and up- $(\uparrow)$ arrows indicate 
quantities of the inner and outer bundles, respectively. 
$\Delta Q(r) \equiv Q_\uparrow - Q_\downarrow$ is a gap of wave number 
at the boundary between the two neighboring bundles. 
In this expression we keep track of the first term only 
and neglect the second term 
because it oscillates much faster than $\theta(x_b)$ changes.

Knowing the above interaction form between the phases 
$\theta_\uparrow$ and $\theta_\downarrow$, 
let us consider a discrete version of the phase Hamiltonian. 
We map the ring crystal onto a simple cubic lattice with size 
$N=L_a \times L_\phi \times L_c$ (See Figure \ref{fig:ring}).
Note that the space is divided uniformly in $\phi$ 
instead of $x_b=r\phi$. 
The lattice spacings are taken as a mesoscopic scale, 
much larger than the CDW wave length.
Adding intra-chain distortion energy and inter-bundle energy on the c-axis 
to eq.(\ref{eq:coupling}), 
the Hamiltonian is written as 
\begin{equation}
H=-\sum_{\alpha=a,b,c}\sum_\mbf{i} 
J_\alpha \cos(\theta_\mbf{i}-\theta_\mbf{i+\hat{\alpha}}
-A_\mbf{i,i+\hat{\alpha}}).
\label{eq:Hamiltonean}
\end{equation}
Here, $\mbf{i}$ labels the lattice points of a simple cubic lattice 
and $\hat{\alpha}$ is the unit lattice vector on the $\alpha$-axis. 
What is important is that there is an additional term, 
\begin{equation}
A_\mbf{i,i+\hat{\alpha}} \equiv \Delta Q x_b \delta_{\alpha a} 
= \Delta N_w \phi \delta_{\alpha a} = 2 \pi f i_b \delta_{\alpha a},
\end{equation}
which comes from the ring geometry. 
$\Delta N_w = r \Delta Q$ is a gap of a number of waves and 
$f=\Delta N_w / L_\phi$ is a filling factor 
that denotes the mean density of vortices. 
$A_\mbf{ij}$ causes a frustration between the couplings along a- and b-axes 
and yields a constant number of phase vortex lines directed to the c-axis 
even without thermal excitation. 
The vorticity is given by the loop integration of $\nabla \theta$ 
along each plaquette. 
By omitting $A_\mbf{ij}$, we obtain a usual XY model, 
which is applicable to CDWs in whisker crystals.

The Hamiltonian eq.(\ref{eq:Hamiltonean}) is equivalent 
to the uniformly frustrated XY model 
which is proposed for the superconductors under magnetic field 
parallel to the c-axis \cite{Li93,Chen97,Hu97}, 
where $A_\mbf{ij}$ is translated to a vector potential in Landau gauge. 
The anisotropy is, however, quite different.
While high $T_c$ superconductor has strong coupling 
inside each CuO$_2$ plane and weak one inter-planes, 
CDW materials has quasi one dimensional property. 
The anisotropy in coupling constants is taken 
as $J_b=\gamma J_a=\gamma J_c, \gamma > 1$ 
in the present model. 
Since the distance between lattice points neighboring on the b-axis 
is proportional to $r$, 
coupling constants and filling factor should depend on the radius $r$ 
but we ignore this $r$-dependence as a first approximation. 

At first, we investigate the ground state of this Hamiltonian. 
By simulated annealing, 
we obtained an energy minimal state for sufficiently large $\gamma$. 
This state has $\Delta N_w$ phase vortex lines 
between every neighboring bc-planes, 
which are straight and parallel to the c-axis. 
These vortex lines form a two dimensional lattice in the ab-plane 
with the unit lattice vectors $(\pm a_0, b_0/2f, 0)$.
Order parameter $r_v$, 
which is related to the Bragg peak height of a vortex lattice, 
can be defined as follows;
\begin{eqnarray}
\nonumber
r_v = S_v(\mbf{q}_v)/N, && \mbf{q}_v = (2\pi /2a_0, 2\pi f/b_0, 0),\\
\nonumber
S_v(\mbf{q}) &=& N^{-1} \langle | v_c(\mbf{q}) |^2 \rangle, \\
v_c(\mbf{q}) &=& \sum_\mbf{j} (\mbf{v_j}\cdot \hat{\mbf{c}}) 
\exp(i\mbf{q \cdot r_j}).
\label{eq:structurefactor}
\end{eqnarray}
Here $(\mbf{v_j}\cdot \hat{\mbf{c}})$ is the vorticity 
defined at the center of a plaquette $\mbf{j}$ 
which is perpendicular to the c-axis.
Owing to the strong coupling along the b-axis, 
the phase is almost uniform along the direction 
but gently modulated with a period $b_0/f$. 
This state looks quite different from the solitonic solution 
in the isotropic case where distortion is localized around the vortex. 
Simple variational calculation supposing the periods 
$2 a_0$ and $b_0/f$ along a- and b-axes, respectively, 
yields an energy minimal solution expressed as 
\begin{equation}
\theta_\mbf{i} = (-1)^{i_a} 
\left[ \frac{\pi}{4} - \frac{2 \gamma ^{-1}}{(2\pi f)^2} 
\cos(2\pi f i_b) \right] 
+ O(\gamma^{-2}),
\label{eq:groundstate}
\end{equation}
which agrees well with the simulated annealing result. 
Note that there are many energetically degenerated states 
which are obtained by the transformation:
$\theta_\mbf{i} \rightarrow \theta_\mbf{i}+2\pi n i_a/ L_a, 
(n=\pm 1, \pm2, \cdots)$. 
This transformation does not change the positions of vortices.


\section{nonequilibrium relaxation analysis of phase transition}
\label{sec:ner}

Next we investigate the relaxation dynamics from an ordered state 
to an equilibrium state at finite temperature, 
i.e., a rapid heating process. 
At the same time, the phase transition is also analyzed 
by the nonequilibrium relaxation method \cite{Ozeki03}.
In order to obtain an initial state, 
we set the phases as eq.(\ref{eq:groundstate}) and 
make the system relax at zero temperature in diffusive dynamics.
After that, Metropolis dynamics at finite temperature is started. 
Filling factor $f$ is fixed to 1/16 in this work. 
The same calculation is done for several anisotropy parameters, 
$\gamma = 10, 16, 24, 32$ and 64. 
The Monte Carlo flip is repeated up to 65,000 steps per each site 
at maximum. 
This step is identified as time, $t$. 
Average is taken over 8-16 samples for each temperature. 
The sample size used is 
$L_a \times L_\phi \times L_c = 64 \times 512 \times 64$ 
and periodic boundary condition is imposed for all directions. 
For this system size, finite size effect 
is not crucial within the observation time.

\begin{figure}
\includegraphics[trim=-0 250 40 -210,scale=0.32,clip]{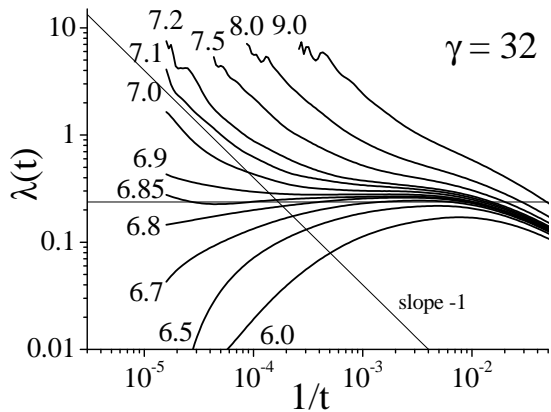}
\caption{\label{fig:local_exp}
The local exponent of the order parameter $r_v$ for various temperatures, 
as a function of $1/t$. 
The temperature of each curve is shown in the figure in the unit of $J_a/k_B$
($T_c=6.87J_a/k_B$). 
The derivative is calculated after local fitting to a 4th order polynomial.
The anisotropy parameter $\gamma$ equals 32. 
The horizontal line shows $\lambda_c$ obtained by dynamical scaling.
}
\end{figure}

The order parameter $r_v$ equals $f^2$ in the initial state 
and decreases to the equilibrium value 
which is zero when $T \ge T_c$  and finite when $T<T_c$. 
To find the critical point, we calculate the local exponent \cite{Ozeki03}, 
\begin{equation}
\lambda(t) = - d \ln r_v(t) / d \ln t.
\end{equation}
In Figure \ref{fig:local_exp}, 
$\lambda(t)$ is plotted with respect to $1/t$.
In the limit of $t \rightarrow \infty$, 
$\lambda(t)$ goes to infinity for $T>T_c$ 
and to zero for $T<T_c$. 
Just at the critical temperature, 
$\lambda(t)$ converges to a certain finite exponent $\lambda_c$
which characterize the critical power law decay.

\begin{figure}
\includegraphics[trim=0 240 40 -210,scale=0.32,clip]{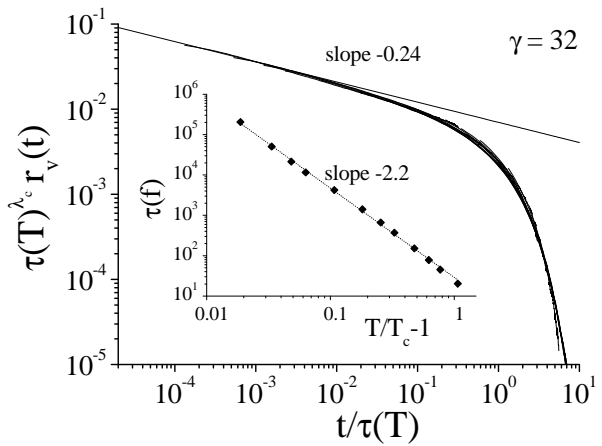}
\caption{\label{fig:scaling}
Scaling plot for $\gamma=32$. 
The temperatures for used data are 
6.9, 7.0, 7.1, 7.2, 7.5, 8.0, 8.5, 9.0, 10.0, 11.0, 12.0 
and 14.0$J_a/k_B$. 
The inset shows critical behavior of the relaxation time.
}
\end{figure}

For $T>T_c$, $r_v$ decays in power function of $t$ for short time scale 
and makes a crossover to exponential decay. 
We can perform dynamical scaling as 
\begin{equation}
r_v(t)=\tau(T)^{-\lambda_c} \tilde{S}_v (t/\tau(T)).
\label{eq:scaling}
\end{equation}
Here $\tau(T)$ is a relaxation time, 
which diverges at the critical temperature $T_c$ as 
\begin{equation}
\tau(T) \propto (T/T_c-1)^{-z\nu}. 
\label{eq:slowingdown}
\end{equation}
We initially obtain $\lambda_c$ and $\tau(T)$'s 
by scaling to eq.(\ref{eq:scaling}) 
then fitting of $\tau(T)$'s to eq.(\ref{eq:slowingdown}) 
yields $T_c$ and $z\nu$.
The scaling result for $\gamma = 32$ is shown in Figure \ref{fig:scaling}. 
The results for several $\gamma$'s are summarized in table \ref{tab:critical}.
$T_c$ is proportional to $(J_b J_c)^{1/2}$, 
which measures the effective elasticity 
of vortex lines confined between two bc-planes.
The exponent $\lambda_c$ seems not universal among different $\gamma$'s. 
It shows, however, a sign for saturation 
to the anisotropic limit value $\approx$ 0.25 as $\gamma$ becomes large.
$z\nu$ is less dependent of $\gamma$ even for small $\gamma$.

\begin{table}[b]
\begin{ruledtabular}
\begin{tabular}{ccccccc}
& $\gamma$ & $T_c$ {\tiny (LE)} & $T_c$ {\tiny (DS)} & $\lambda$ {\tiny (DS)} & $z\nu$ {\tiny (DS)} &\\
\hline
&10 & 1.159(2) & 1.13 & 0.06 & 2.2 &\\
&16 & 1.190(5)  & 1.17 & 0.16 & 2.1 &\\
&24 & 1.208(6)  & 1.18 & 0.22 & 2.2 &\\
&32 & 1.214(5)  & 1.20 & 0.24 & 2.2 &\\
&64 & 1.206(6)  & 1.18 & 0.25 & 2.6 &\\
\end{tabular}
\end{ruledtabular}
\caption{\label{tab:critical}
Critical temperature in the unit of $(J_bJ_c)^{1/2}/k_B$ 
and exponents for several anisotropy parameters. 
LE means the critical temperature is estimated 
from the long time behavior of local exponent 
and DS means dynamical scaling. 
}
\end{table}

\begin{figure}
\includegraphics[trim=0 250 40 -210,scale=0.32,clip]{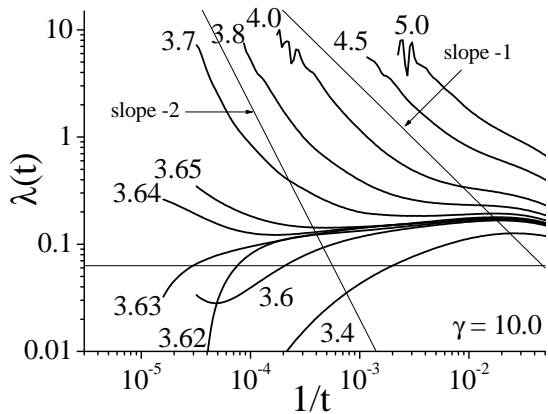}
\caption{\label{fig:local_exp_j10}
The local exponent for a relatively small anisotropic parameter $\gamma=10$. 
}
\end{figure}

For small $\gamma$ dynamical scaling does not work well 
{\it in the very vicinity of} the critical temperature,  
where the functional form of relaxation changes with temperature. 
In Figure \ref{fig:local_exp_j10}, the local exponent for $\gamma=10$ is 
plotted with $1/t$. 
If the relaxation form is simple exponential one in the long time limit, 
$\lambda(t)$ would be asymptotically proportional to $t$. 
Figure \ref{fig:local_exp_j10} shows, however, $\lambda(t) \propto t^{\mu}$ 
where $\mu$ is the larger than unity 
and increases up to 2 as approaching the critical temperature. 
Such bad scaling region in temperature, 
however, becomes narrower as $\gamma$ increases.

\section{relaxation dynamics in ordered phase}

\begin{figure}
\includegraphics[trim=0 250 40 -210,scale=0.32,clip]{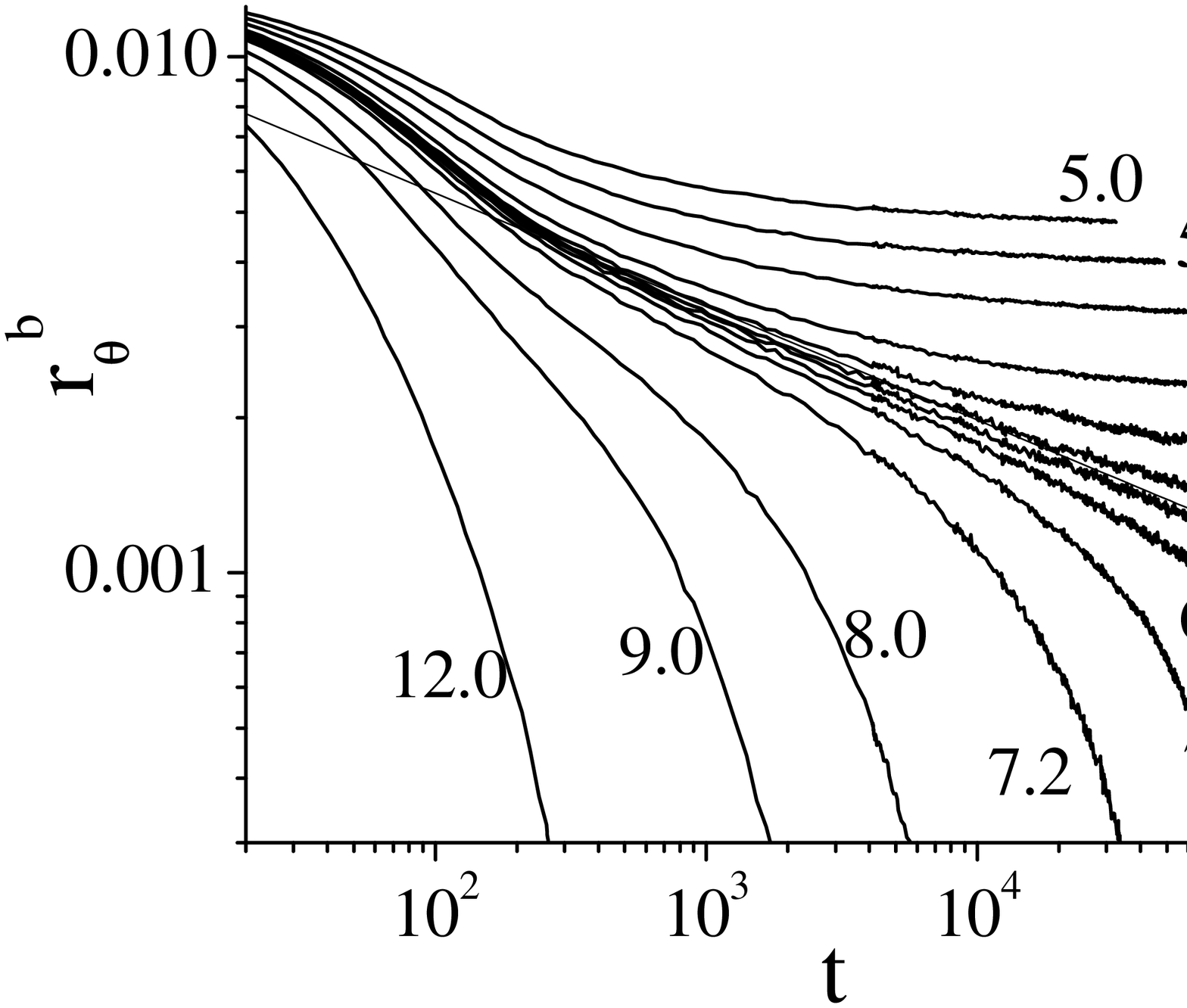}
\includegraphics[trim=0 250 40 -210,scale=0.32,clip]{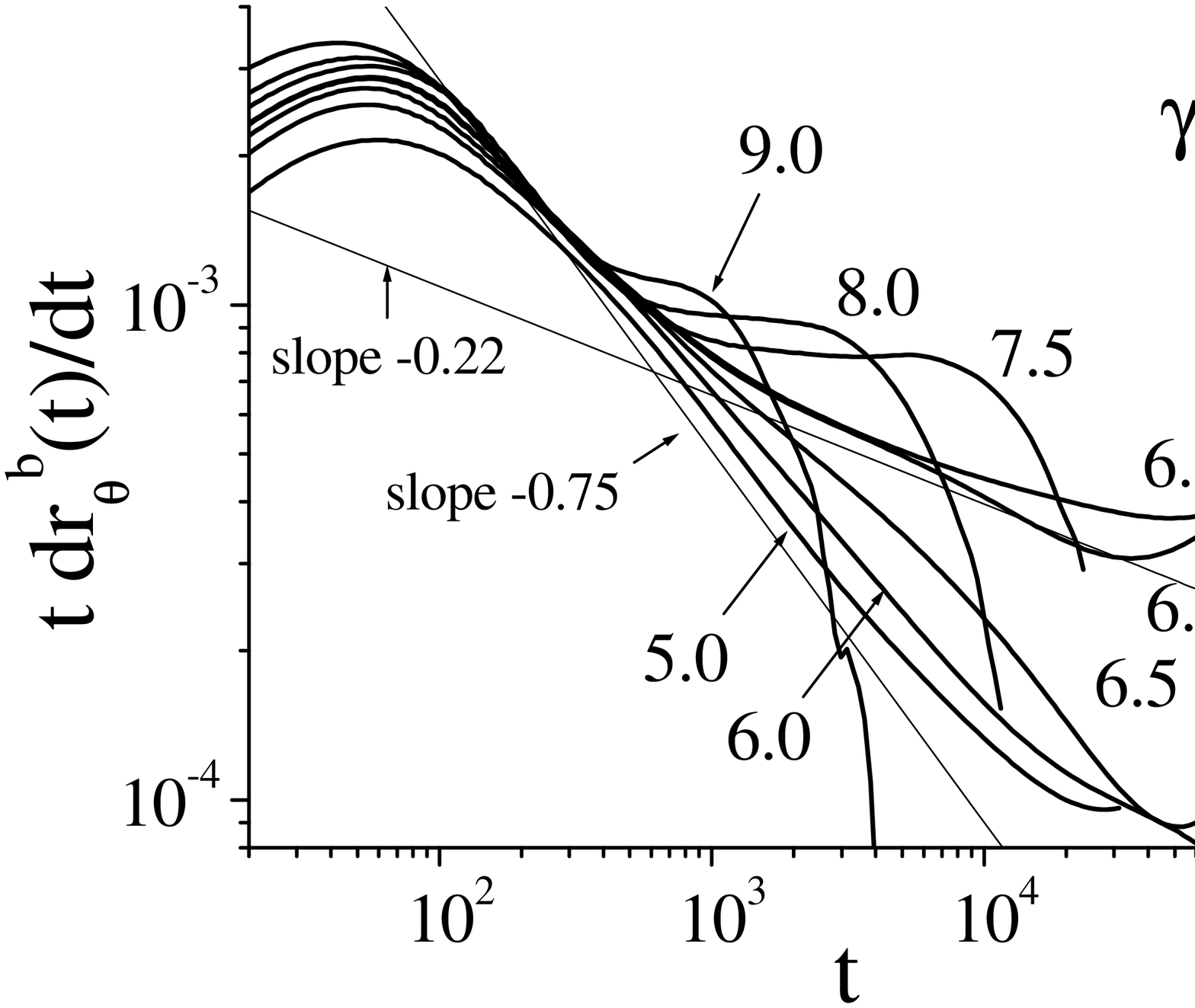}
\caption{\label{fig:drxdt-t}
Time evolution of phase structure factor $r_\theta^b$ (top) 
and time derivative of $r_\theta^b$ {\it multiplied by $t$} (bottom). 
The derivative is obtained 
by local fitting to a 4th order cubic polynomial. 
The line of slope -0.22 means the relaxation 
at the critical temperature.
}
\end{figure}

Next, we investigate the relaxation function in the low temperature phase. 
In order to eliminate the influence of the finite equilibrium value, 
the time derivative of the order parameter is calculated. 
The relaxation of $r_v$ does not look like an exponential function 
in the time range of the present simulations. 
We don't find an apparent reduction of relaxation time 
when leaving from the critical point unlike in the high temperature phase. 
Although the functional form of $r_v$ is unclear, 
the power law behavior is observed to be suitable for some other quantities, 
e.g., total energy of the system 
and the order parameter of phase periodicity $r_\theta^b$.
Here 
\begin{equation}
r_\theta^b \equiv S_\theta(\mbf{q}_{\theta}^b)/N. 
\end{equation}
Here $S_\theta(\mbf{q})$ is a Fourier transformation of 
phase correlation function defined as 
\begin{equation}
S_\theta(\mbf{q})=N^{-1} \sum_\mbf{i,j} 
\cos(\theta_\mbf{i}-\theta_\mbf{j}) 
\exp \left[i \mbf{q \cdot (r_i-r_j) }\right], 
\end{equation}
which has peaks at $\mbf{q}_{\theta}^a \equiv (2\pi /a_0, 0,0)$
and $\mbf{q}_{\theta}^b \equiv (0,2\pi f/b_0,0)$.
Note that this order parameter is finite at $t=0$ 
only because the initial state is chosen as eq.(\ref{eq:groundstate}) 
and other metastable states have peaks in different wave number vectors. 
The time evolution of $r_\theta^b$ 
and $t d r_\theta^b/dt$ is shown in Figure \ref{fig:drxdt-t}.
For $T<T_c$, $r_\theta^b$ seems to decay to a finite value 
in the limit of $t \rightarrow \infty$. 
This indicates that not only vortex but also phase has 
true long range periodic order in three dimensions. 
We confirm that this finite value does not come from the finite size effect. 
The time derivative shows power law relaxation with $t$. 
The decay exponent becomes larger with $T$.

\section{discussion}

\subsection{dimensionality of periodic order}

So far, it is shown that the quasi one dimensional frustrated XY model 
takes a vortex lattice melting transition. 
The dynamical scaling analysis strongly suggests  
that it is a standard second order transition. 
Then $r_v$ and $r_\theta^b$ work as order parameters 
of vortex and phase itself.

By equilibrium simulations, we find that helicity modulus, which measures 
the thermodynamic stiffness of phase against the perturbation, 
$\theta_\mbf{i} \rightarrow \theta_\mbf{i} + \epsilon i_\alpha 
(\epsilon \rightarrow 0)$
\cite{Li93}, 
is zero along the a-axis for all temperature 
while those along b- and c-axes are finite below $T_c$. 
This means that the (quasi) long range phase coherence 
is established only in the ab-plane. 

The reason why the two dimensional phase coherence appears 
in quasi one dimensional system is considered as follows: 
Even with $J_a=J_c (\ll J_b)$, 
the a- and c-axes are not equivalent 
since the former conflicts with the b-axis and the latter does not. 
The fast growth of the order along the b-axis suppresses 
the order along the a-axis 
then two dimensional-like order appears.

Power law behavior and two dimensional order remind us of 
the Kosterlitz-Thouless (KT) phase \cite{Kosterlitz73} 
with quasi long range order. 
If so, it is strange that $r_\theta^b$ is finite in the long time limit 
as shown in the previous section. 
Furthermore finite $r_\theta^b$ seems directly inconsistent 
with the fact that helicity modulus for the a-axis is zero. 
Of course we can not eliminate the possibility 
that $r_\theta^b$ decreases to zero in the time scale 
much longer than the present observation time 
but there can be another explanation as follows.  
When the two dimensional {\it true} long range order is established, 
the center of mass of its phase is individually pinned in every bc-plane 
due to the break of ergodicity. 
As a result the phase fluctuation along the a-axis is suppressed 
even though there is no restoring force. 
Thus finite $r_v$ which is owing to the initial condition survives 
even though long range order exists only in two dimensions. 
In contrast, vorticity is a gauge invariant quantity 
and really has three dimensional long range order. 
If this is true, 
there would be another mechanism for the power decay 
which differs from the KT scenario, 
on which $r_\theta^b$ goes to zero. 
Indeed, power divergence of relaxation time at $T_c$ is 
inconsistent with the exponential divergence 
for the KT transition \cite{Ozeki03}.

\subsection{comparison with in-plane isotropic case}

Here we compare the present result with high $T_c$ superconductors.
The case of the magnetic field perpendicular to the CuO$_2$ plane 
is given by $J_a = J_b \gg J_c$ in the present notation. 
This is isotropic in the frustrated ab-plane. 
It is believed that there is a first order transition \cite{Hu97} 
into the low temperature phase 
where phase coherence exists only along the c-axis \cite{Chen97}. 
The limit $\gamma \rightarrow 1$ in the present model 
is also expected to show the same transition 
although the discontinuous property is quite small 
in the case that $J_c$ is as large as $J_a,J_b$ \cite{Chen97}. 
There is a possibility that bad scaling behavior 
near $T_c$ for smaller $\gamma$ as mentioned 
at the last of section \ref{sec:ner} 
is due to the crossover to a first order transition.

Power law relaxation is also observed 
in the isotropic frustration model \cite{Li93}. 
This is apparently unrelated to the KT order. 
We consider that the power law behaviors 
of the isotropic and anisotropic models 
is a common property of the partially ordered states 
of frustrated systems 
although dimensionality of phase ordering is different.

\subsection{comparison with quasi two dimensional case}

The case of a magnetic field parallel to the CuO$_2$ plane 
can be expressed by the present model with $\gamma \ll 1$ 
although the CuO$_2$ plane is normal to the b-axis, not to the c-axis 
\cite{Balents94}. 
This is just the opposite anisotropy to the present study 
and the system is divided into isolated planes without frustration 
in the strong anisotropy limit. 
This case, however, has many common properties 
with the present case of $\gamma \gg 1$ 
in spite of the difference in quasi two and one dimensional properties, 
e.g., two dimensional phase coherence parallel to the c-axis 
and the same structure of vortex lattices \cite{Hu00,Hu04}. 
Additionally 
the phase transitions change from first order to second order 
as anisotropy becomes stronger in both cases. 
This similarity comes from the fact 
that qualititative behavior is not affected 
by the strength of the coupling along the c-axis, 
which is not concerned with the frustration of the system. 
It is essential that both of these opposite anisotropies, 
$\gamma \gg 1$ and $\gamma \ll 1$,  
break the balance of frustration in the ab-plane in the same way. 

The vortex lattice order in the quasi two dimensional system 
is considered to be of quasi long range \cite{Hu04}. 
It is also reported that the two layer system 
(this is purely two dimensional) 
has two types of KT phases \cite{Orignac01, Granato05}, 
the vortex phase and the Meissner phase. 
It is, however, not clear whether 
the inter layer coupling in three dimensional system 
could be irrelevant or not for small but finite $\gamma$. 
We consider that true long range order 
assisted by three dimensional vortex lattice order 
is worth to re-examine 
in quasi two dimensional case as well as the present case. 


\subsection{Experiment}

Finally let us mention about the relation to the experiments 
of a CDW in a ring crystal\cite{Shimatake06}. 
The present work based on the anisotropic frustrated XY model 
found a phase transition to the ordered state 
where phase vortex lines along the c-axis form a lattice 
and two dimensional phase coherence is established 
in each cylindrical shell perpendicular to radius direction. 
Thus it suggests 
that the ring geometry makes the CDW phase transition 
quite different from that of a whisker crystal. 
The latter is described by the XY model without frustration, 
which shows a simple ferromagnetic transition. 
In the low temperature phase of a ring crystal, 
the relaxation of phase fluctuation, 
which is closely connected to electric polarization and reflectivity, 
shows a power law relaxation without characteristic time scale, 
so that small apparent relaxation time, 
which does not show singular behavior, would be estimated 
if one supposes an exponential decay as for whisker crystals. 

There remain some future works.
In this work, rapid heating process is studied 
because of its self averaging property instead of rapid cooling process.
The latter is a challenging work 
because it is proper for the direct comparison 
with the laser pumping experiment 
and contains the process to untwist entangled flux lines. 
The present model drops some features of the system, 
e.g., radius dependence of the model parameters, 
which causes distribution of local critical temperature 
and relaxation time then the transition could be blurred,  
and a pinning effect of lattice defects in atomic crystal.

The authors thank S. Tanda, Y. Toda and K. Shimatake 
for useful discussions. 
This work is supported by 21st Century COE program 
``Topological Science and Technology''.



\end{document}